\documentclass{iau} 
\usepackage{graphicx}

\title[UVIT observations] 
{Understanding Galaxy Mergers \& AGN Feedback with UVIT}

\author[Rubinur K., M. Das, P. Kharb, \& Rahne P. T]   
{K. Rubinur$^1$,
M. Das$^2$, P. Kharb$^1$, \and Rahne P. T$^2$}

\affiliation{
$^1$ National Centre for Radio Astrophysics - Tata Institute of Fundamental Research, S. P. Pune University Campus, Ganeshkhind, Pune, 411007, India, email: {\tt rubinur@ncra.tifr.res.in}\\
$^2$ Indian Institute of Astrophysics, 2nd Block, Koramangala, Bengaluru, 560034, India.}

\pubyear{2019}
\volume{356}  
\jname{Nuclear Activity in Galaxies Across Cosmic Time}
\editors{A.C. Editor, B.D. Editor \& C.E. Editor, eds.}
\begin{document}

\maketitle

\begin{abstract}
Simulations expect an enhanced star-formation and active galactic nuclei (AGN) activity during galaxy mergers, which can lead to formation of binary/dual AGN. AGN feedback can enhance or suppress star-formation. We have carried out a pilot study of a sample of $\sim$10 dual nuclei galaxies with AstroSat's Ultraviolet Imaging Telescope (UVIT). Here, we present the initial results for two sample galaxies (Mrk~739, ESO~509) and deep multi-wavelength data of another galaxy (Mrk~212). UVIT observations have revealed signatures of positive AGN feedback in Mrk~739 and Mrk~212, and negative feedback in ESO~509. Deeper UVIT observations have recently been approved; these will provide better constraints on star-formation as well as AGN feedback in these systems.
\end{abstract}

\firstsection 
\section{Introduction}

Galaxies evolve through major and minor mergers as well as through interaction with nearby galaxies (Barnes \& Hernquist 1992). 
Simulations show that galaxy mergers/interactions cause gas inflow onto the nuclei resulting in the accumulation of dense gas around the nuclei (Bournaud et al. 2010). This triggers vigorous star-formation around the single or double nuclei. Hence, mergers and close interactions of galaxies often fall in the category of starburst galaxies (Bournaud et al. 2010). Models of star-formation predict an increase of the star formation rate both in the disks, nuclei and even in the outer tidal tails (e.g. Duc et al. 2000). However, it has been found that the increase in star-formation is much lower than what is considered a typical starburst (Ellison et al. 2013). 
As the nuclear mass concentration and gas inflow increases, active galactic nuclei (AGN) activity may be triggered in one or both nuclei. This is due to mass accretion onto the supermassive black holes (SMBHs) of the individual galaxies which can lead to binary/dual AGN (Begelman et al. 1980). Studying dual AGN is one of the ways in which we can follow the SMBHs as they sink into the centres of the merger remnants and finally coalesce in the central bulge. There are several studies which have attempted to detect these binary/dual AGN (Rubinur et al. 2017, 2019 and refs. therein).

Once AGN activity is triggered and the SMBHs reach a certain critical mass (Ishibashi \& Fabian 2012), they give out energy to the surrounding medium via winds, jets and radiation. This can enrich the circum-galactic medium (CGM). The winds also trigger star formation beyond the AGN by shocking gas and suppress gas infall by blowing out the gas. This is collectively called AGN feedback (See Harrison et al. 2017 for review). UV emission is a good tracer of the star formation rate (SFR). It can trace recent as well as older star formation. High resolution UV observations can help us understand how SFR is affected by mergers and AGN feedback. Hence, the new ultra-violet telescope UVIT on the Astrosat satellite (Kumar et al. 2012) with a spatial resolution of 1.2$^{\prime\prime}$ is crucial for tracing the star formation rates (SFRs) in nearby galaxies.
We have carried out a pilot study of dual nuclei galaxies with UVIT to detect merger- and AGN- related star-formation in galactic disks and tidal tails. Here, dual nuclei can be a pair of AGN, AGN-starburst nuclei or starburst-starburst. We have selected our sample from surveys of dual core galaxies (e.g. Mezcua et al. 2014) and confirmed dual AGN or AGN-starburst nuclei (Koss et al. 2012).
We had chosen 17 nearby galaxies with redshifts $<0.1$, having dual nuclei with projected separations below 10 kpc and exhibiting strong UV emission in GALEX observations.

So far, we have observed $\sim$10 sources with UVIT using short exposures ($<$3~ks). 
Multi-frequency observations have also been carried out for one galaxy. Here, we present results from the initial observations of two galaxies and multi-wavelength data along with deep UVIT observation of Mrk~212 (Rubinur et al., 2019, MNRAS, submitted).   

\section{Results:}
\noindent
{\underline{\it Mrk~739:}} is a confirmed dual AGN (Koss et al. 2011). One of the nuclei is a Seyfert type 1 while the other is a Seyfert type 2.
The nuclei are at a projected separation of 3 kpc. The optical spectra show signatures of outflows from one of the nuclei, while the molecular gas profile is asymmetric (Koss et al. 2011). Our UVIT images (2.5 ks) show the star forming regions around the nuclei of Mrk~739 (Fig 1). These star-forming knots can be induced by the positive AGN feedback activity.  Deeper UV observations are required to confirm direct signatures of AGN-feedback-related star-formation.

\noindent
{\underline{\it ESO 509-IG 066 NED 02 (ESO 509):}} is a confirmed dual AGN in an ongoing galaxy merger; the two nuclei are at a projected separation of 11.2~kpc (Koss et al. 2012). 
Our 2.5 ks UVIT image shows UV emission from the individual galaxies. Furthermore, we have found that three UV holes (Fig 1) in Source 2, surrounding the central nuclei. These could be signatures of negative AGN feedback (e.g. George K. et al. 2019).

\begin{figure}[b]
\begin{center}
 \includegraphics[width=2.6in]{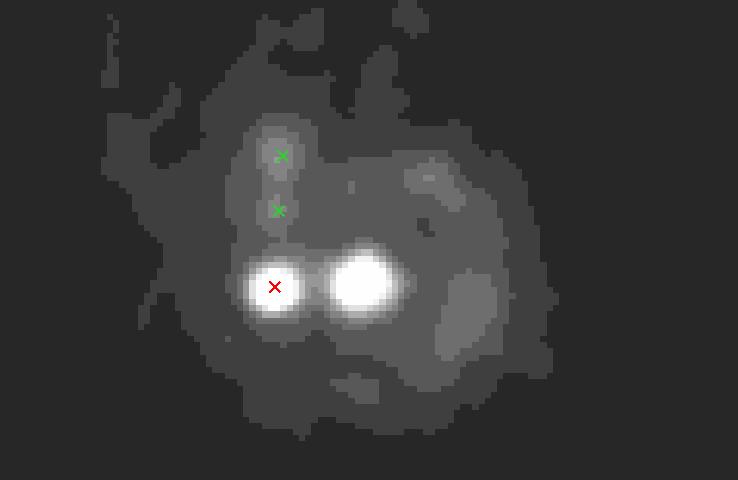} \includegraphics[width=2.6in]{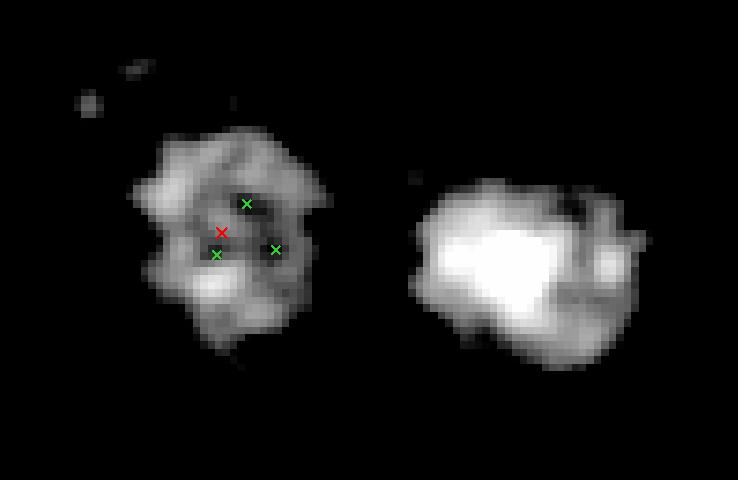}
 \caption{(left) UVIT image of Mrk 739. It shows UV emission from two nuclei and outer disk. Two star-forming knots (green cross) are detected, which may be related to one of the nuclei (red cross). (right) UVIT image of ESO 509. It shows three cavities (green cross). Red cross is the centre of the galaxy.}
   \label{fig1}
\end{center}
\end{figure}

\begin{figure}
\begin{center}
\includegraphics[width=3.1in]{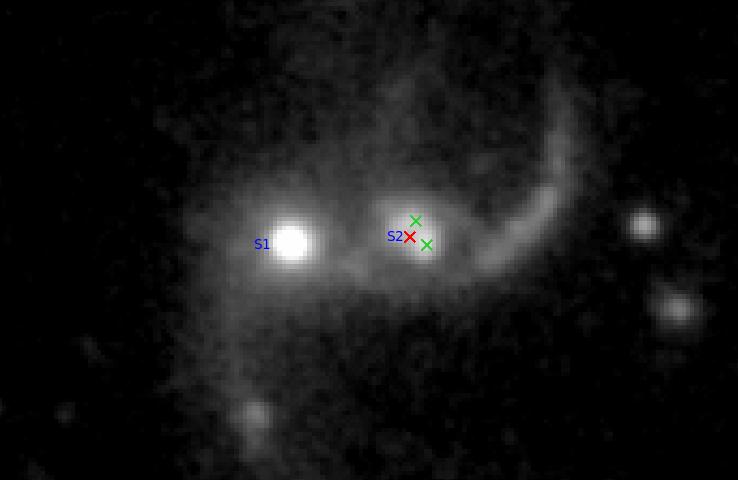} 
\caption{This is a deep UVIT image of MRK 212 (15 ks). It has resolved star-forming knots in S2 (green cross) and tidal arms. The red cross is the centre of S2.}
  \label{fig1}
\end{center}
\end{figure}

\noindent
{\underline{\it Mrk~212:}}
is a double-pinwheel galaxy with a companion. 
Mrk 212 has two radio sources (S1 and S2) with a projected separation of $\sim$6 kpc (Fig 2).
The 15 GHz VLA image and $1.4-8.5$~GHz core spectral index ($-0.81\pm0.06$) supports the presence of an AGN in S1. S2 has a compact structure at 15 GHz which coincides with the optical centre of the companion galaxy and an extended structure at 8.5 GHz, which is offset by $\sim$ 1$^{\prime\prime}$ from the optical centre. Harnazndez-Ibarra et al. (2016) have identified S2 as an AGN, from optical spectroscopy. Our deep UVIT observations of Mrk 212 resolve the star-forming knots in S2 (Fig 2) and detect tidal tails. The star-forming knots in S2 coincide with the two sided radio structure detected at 8.5 GHz, which could be the result of positive AGN feedback in S2. However, the overall SFR is similar to isolated galaxies. This is not typically expected in the standard picture of galaxy merger and subsequent enhanced SFR.


\section{Summary}
We have carried out UVIT observations of $\sim$10 dual nuclei galaxies. These observations have revealed signatures of AGN feedback in Mrk~739, ESO~509 and Mrk~212. While Mrk 739 and Mrk 212 may show positive AGN feedback, ESO 509 shows negative AGN feedback. We require high sensitivity radio and molecular gas observations in Mrk 212 to confirm this feedback. Mrk 739 and ESO 509 will be observed for 15 ksec in the upcoming UVIT cycle, which would confirm the star-forming knots as well as UV cavities in these two systems. While some studies find that mergers can enhance  star-formation, others have shown the opposite trend (See Pearson et al. 2019). We find that Mrk~212 has an SFR similar to isolated galaxies, consistent with the latter. 
Systematic multi-wavelength observations of large samples of galaxies are required to fully understand AGN feedback.


\begin{thebibliography}{}

\bibitem[{{Barnes} \& {Hernquist}(1992)}]{barnes1992}
{Barnes} J.~E., {Hernquist} L., 1992, \textit{araa}, 30, 705

\bibitem[{Bournaud}]{Bournaud2010}
{Bournaud} F., 2010, \textit{arXiv0909.1812}

\bibitem[{Duc, P.-A.; Brinks, E.; Springel, V.}]{duc2000}
Duc, P.-A.; Brinks, E.; Springel, V.; et al. 2000, \textit{AJ}, 120, 1238

\bibitem[Ellison]{elisson2013}
Ellison, S. L., Mendel, J. T., Patton, D. R., \& Scudder, J. M. 2013, \textit{MNRAS}, 435, 3627

\bibitem[Begelman]{Begelma1980}
Begelman M. C., Blandford R. D., Rees M. J., 1980, \textit{Nature}, 287, 307

\bibitem[Ishibashi]{Ishibashi2012}
Ishibashi W., Fabian A. C., 2012, \textit{MNRAS}, 427, 2998

\bibitem[Harrison]{Harrison2017}
Harrison C. M., 2017, \textit{Nature Astronomy}, 1, 0165

\bibitem[Kumar]{Kumar2012}
Kumar A. et al., 2012, \textit{SPIE Conference Series}, Vol. 8443,
procspie, p. 84431N

\bibitem[Koss]{koss}
Koss, M; Mushotzky, R; Treister, E; et al.  2012, \textit{ApJ}, 746L, 22

\bibitem[Mezcua]{Mezcua2014}
Mezcua, M.; Lobanov, A. P.; Mediavilla, E.; Karouzos, M. 2014, ApJ, 784, 16

\bibitem[Hernandez]{Hernandez}
Hernandez Ibarra F. et al., 2016, MNRAS, 459, 291

\bibitem[W. J. Pearson]{W. J. Pearson2019}
Pearson, W. J. et al. 2019 \textit{AA},  631, A51

\bibitem[rubunur]{rubinur2017}
Rubinur K., Das M., Kharb P., Honey M., 2017, \textit{MNRAS},
465, 4772


\bibitem[rubunur]{rubinur2019}
Rubinur K., Das M., Kharb P., 2019, \textit{MNRAS}, 484, 4933



\bibitem[george]{george2019}
George K. et al. 2019, \textit{MNRAS} 487,3102
\end{thebibliography}

\end{document}